\pdfoutput=1
\RequirePackage{fix-cm}
\documentclass[onecolumn]{svjour3}                     % onecolumn (standard format)
\smartqed  % flush right qed marks, e.g. at end of proof
\usepackage{graphicx}
\usepackage{bm}
\usepackage{amsmath}
\usepackage{amssymb}
\usepackage{mathptmx}      % use Times fonts if available on your TeX system
\usepackage{latexsym}
\usepackage{verbatim}
\usepackage{color}
\usepackage[english]{babel}
\usepackage{hyperref}

\journalname{Journal of Computational Electronics}
\begin{document}
%single-molecule nanojunctions
\title{First-principles vs. semi-empirical modeling of global and local electronic transport properties of graphene nanopore-based sensors for DNA sequencing}


\titlerunning{Total and local electronic currents in graphene nanopores for DNA sequencing}        % if too long for running head

\author{Po-Hao Chang \and Haiying Liu \and
        Branislav K. Nikoli\' c
}


\institute{Po-Hao Chang \and
           Branislav K. Nikoli\' c  \at
           Department of Physics and Astronomy, University of Delaware, Newark, DE 19716, USA  \\
              \email{bnikolic@udel.edu}           %  \\
           \and
           Haiying Liu
            \at
            Department of Physics and Astronomy, University of Delaware, Newark, DE 19716, USA  \\
            School of Physics and Technology, University of Jinan, Jinan 250022, PR China
}

\date{Received: date / Accepted: date}

\maketitle

\begin{abstract}
Using first-principles quantum transport simulations, based on the nonequilibrium Green function formalism combined with density functional theory (NEGF+DFT), we examine changes in the total and local electronic currents within the plane of graphene nanoribbon with zigzag edges (ZGNR) hosting 
a nanopore which are induced by inserting a DNA nucleobase into the pore. We find a sizable change of the zero-bias conductance of two-terminal ZGNR + nanopore device after the nucleobase is placed into  the most probable position (according to molecular dynamics trajectories) inside the nanopore of a small diameter \mbox{$D=1.2$ nm}. Although such effect decreases as the nanopore size is increased to \mbox{$D=1.7$ nm}, the contrast between currents in ZGNR + nanopore and ZGNR + nanopore + nucleobase systems can be  enhanced by applying a small bias voltage $V_b \lesssim 0.1$ V. This is explained microscopically as being due to DNA nucleobase-induced  modification of spatial profile of local current density around the edges of ZGNR. We repeat 
the same analysis using NEGF combined with self-consistent charge density functional tight-binding (NEGF+SCC-DFTB) or self-consistent extended  H\"{u}ckel (NEGF+SC-EH) semi-empirical methodologies. The large discrepancy we find between the results obtained from NEGF+DFT vs. those obtained from NEGF+SCC-DFTB or NEGF+SC-EH approaches could be of great importance when selecting proper computational algorithms for {\em in silico} design of optimal nanoelectronic sensors for rapid DNA sequencing.

\end{abstract}

\section{Introduction}\label{sec:intro}

The successful realization of fast and low-cost methods for reading the sequence of DNA nucleobases is envisaged to lead to personalized medicine and applications in various subfields of genetics~\cite{Schadt2010}. The use of nanometer-sized pores provides a simple idea that can lower the cost and speed up DNA sequencing by eliminating enzyme-dependent amplification and fluorescent labeling steps. Two major types of nanopores~\cite{Venkatesan2011,Wanunu2012}  have been employed for the so-called third generation DNA sequencing~\cite{Schadt2010}: ({\em i}) protein nanopores (such as $\alpha$-hemolysin pore); and ({\em ii}) artificial solid-state pores~\cite{Dekker2007} (such as silicon nitride and silicon oxide pores). In these schemes, DNA molecules in electrolytic solution are electrophoretically driven through the nanopore, and one tries to detect 
the sequence of nucleobases by monitoring how they  reduce longitudinal ionic current flowing through the nanopore~\cite{Dekker2007}.

{\em The key issues in these approaches are how to slow down the translocation speed of DNA and how to achieve single-base resolution}. In particular, solid-state nanopores~\cite{Venkatesan2011,Wanunu2012} represent an inexpensive and highly versatile alternative to initially considered biological nanopores since they provide superior mechanical, chemical and thermal characteristics when compared with lipid-based systems. At the same time, the size and shape of the nanopore can be tuned with sub-nanometer precision; high-density arrays of nanopores can be easily fabricated~\cite{McNally2010}; and they can be integrating with electronic~\cite{Merchant2010,Schneider2010,Garaj2010}  or optical~\cite{McNally2010} readout techniques. Despite much progress made in nanopore sequencing techniques, it is still difficult to resolve nucleotides at the level of single-base resolution because the conventional nanopores are several nanometers in length so that 10--15 nucleotides occupy them at a given time.

Very recent experiments~\cite{Merchant2010,Schneider2010,Garaj2010,Traversi2013} on DNA translocation through graphene nanopores have introduced a new contender into this arena. Graphene~\cite{Geim2009}---a two-dimensional allotrope of carbon whose atoms are densely packed into a honeycomb lattice---brings its unique mechanical and electronic properties into the search for an optimal nanoelectronic sensor for rapid DNA sequencing. Most importantly, single layer graphene is only one-atom-thick so that the entire thickness of the nanopore through which single-stranded DNA (ssDNA) is threaded is comparable to the dimensions of DNA nucleotides (e.g., the spacing between nucleotides in ssDNA is 0.32--0.52 nm, while the ``thickness'' of single layer graphene is 0.34 nm). 

However, the recent experiments~\cite{Merchant2010,Schneider2010,Garaj2010} on graphene nanopores, which have measured fluctuations in the vertical ionic current flow due to double-stranded DNA (dsDNA) translocation through the pore,  have not reached sufficient resolution to detect and identify individual nucleobases. This is mainly due to: ({\em i}) high DNA translocation velocity in graphene nanopores ($> 40$ nucleotides/$\mu$s) pushes the detector bandwidth requirements to the MHz region, which precludes the measurement of pA steps in ionic current; and ({\em ii})  high $1/f$ noise in graphene nanopores can reduce the detector signal-to-noise ratio and potentially prohibit the direct measurement of individual nucleotides using ionic current. Another critical issue is that DNA easily sticks to graphene due to its hydrophobicity which can clog the nanopore and prevent DNA translocation, as encountered in the recent experiments~\cite{Schneider2013} and explained by all-atom molecular dynamics (MD) simulations~\cite{Wells2012}. Nevertheless, the recent functionalization of graphene surface with a self-assembled monolayers (such as pyrene ethylene glycol) has demonstrated how to make graphene surface hydrophilic, where such monolayer is not bonded covalently to graphene in order to avoid strong modification of its electronic  properties~\cite{Schneider2013}.

\begin{figure*}
\begin{center}
\includegraphics[scale=0.37,angle=0]{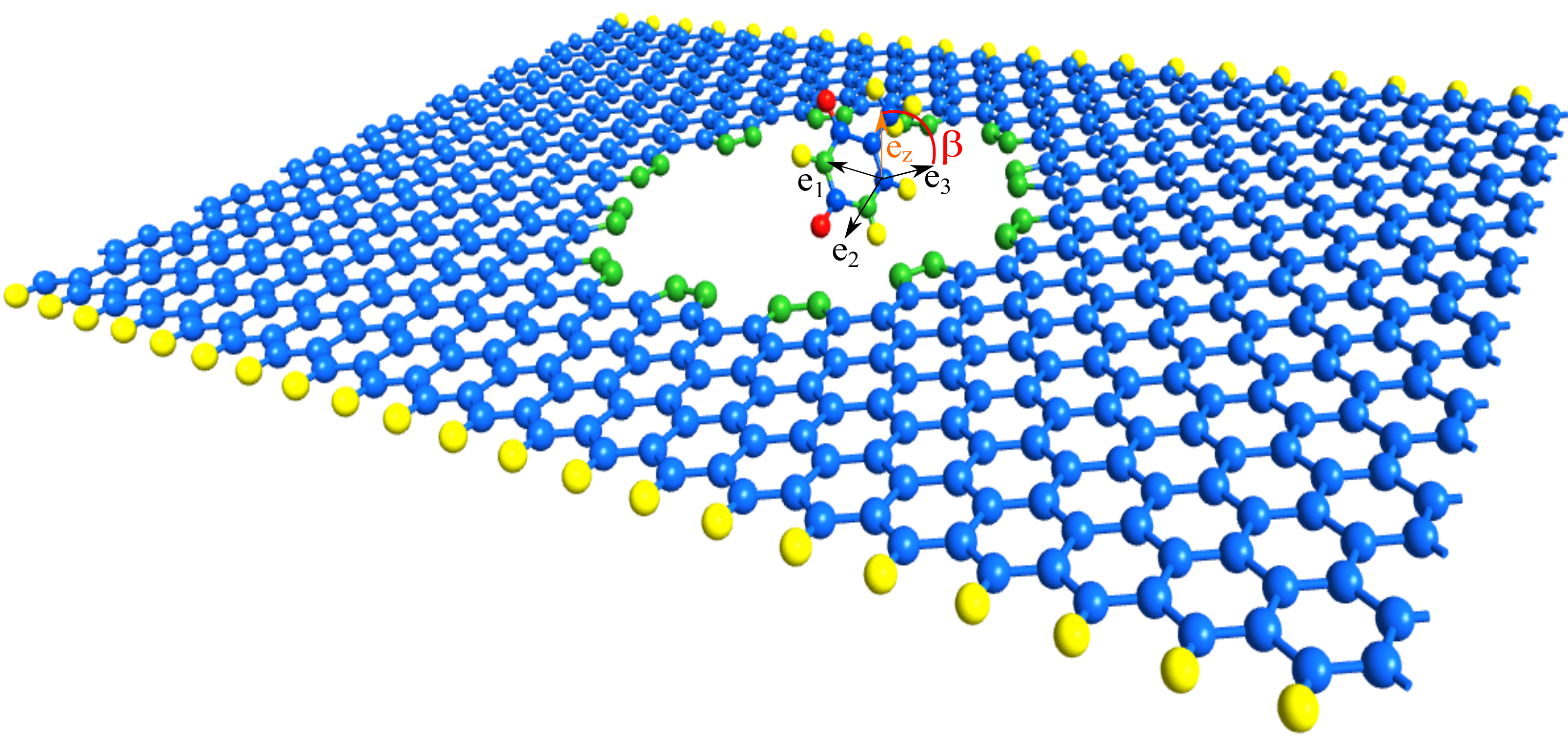}
\end{center}
\caption{Schematic view of the two-terminal sensor for DNA sequencing where transverse electronic current flows in the plane of metallic GNR with zigzag edges hosting a nanopore through which the DNA molecule is translocated longitudinally. The active device region consists of a segment of 14-ZGNR or 16-ZGNR hosting nanopore of diameter $D=1.2$ nm or $D=1.7$ nm, respectively. The edge carbon atoms of the nanopores are assumed to be passivated by nitrogen, while edge carbon atoms of ZGNR itself are passivated by hydrogen. The T nucleobase is inserted into the nanopore in the most probable position according to MD trajectories of Ref.~\cite{Wells2012}, with distance between the nucleobase center of mass and nanopore wall being \mbox{$\simeq 1.5$ \AA} and tilt angle $\beta \simeq 68^\circ$. The total number of simulated atoms (C-blue, H-yellow, N-green, O-red) in the active region, including T nucleobase within the nanopore, is $657$ in the case of 16-ZGNR and $515$ in the case of 14-ZGNR (Color figure online)}
\label{fig:fig1}
\end{figure*}

Graphene also offers unique properties~\cite{DasSarma2011} of electronic current flowing in the plane due to its charge carriers behaving as massless Dirac fermions. However, current flowing through large-area graphene with a nanopore is largely unaffected by the presence of DNA nucleobases within the nanopore~\cite{Rocha2013}. Following theoretical proposals~\cite{Zwolak2008,Lagerqvist2006,Krems2009,Meunier2008,Chen2012} and experiments~\cite{Tsutsui2010,Huang2010} on nanogaps between two metallic electrodes (such as gold~\cite{Zwolak2008,Lagerqvist2006,Krems2009} or CNTs~\cite{Meunier2008,Chen2012}), where the longitudinally translocated DNA through the gap modulates the transverse tunneling current across the gap, a number of recent studies have analyzed potential of tunneling current across a nanogap~\cite{Postma2010,Prasongkit2011,He2011} between two metallic graphene nanoribbons (GNRs) or a nanopore~\cite{Nelson2010,Girdhar2013,Qiu2014} within semiconducting GNRs to detect nucleobases. In the case of nanogaps between gold electrodes, the tunneling currents are off-resonant, reaching pA at typically applied~\cite{Tsutsui2010,Huang2010} bias voltage \mbox{$\simeq 0.5$ V} and being highly dependent on difficult-to-control relative geometry between the molecule and electrodes. Thus, recent experiments~\cite{Tsutsui2010,Huang2010,Chang2010} have focused on extracting signal from noise via intricate statistical analysis~\cite{Huang2010,Lagerqvist2006,Krems2009} aiming to identify nucleobases by repeating thousands of measurements. The tunneling current signal in the case of GNR electrodes is further reduced due to much smaller number of available conducting channels and weaker coupling between graphene electrodes and DNA (because of much smaller spatial extension of carbon outer orbitals compared to those of gold).

The very recent experiments~\cite{Traversi2013}, conducted on GNRs of width \mbox{$W \sim 100$ nm} hosting a nanopore of diameter \mbox{$D \sim 10$ nm}, have detected DNA translocation through the nanopore by observing \mbox{$\simeq 10$\%} change in the in-plane electronic current (of the order of \mbox{100 nA} driven by the source-drain bias voltage of \mbox{$\simeq 20$ mV}). However, poor signal-to-noise ratio prevents this sensor, or the other GNR-based concepts discussed above, to perform full DNA sequencing operation without involving complex data analysis (such as using cross-correlations of currents due to the same nucleobase passing sequentially through several nanopores in multilayered graphene~\cite{Ahmed2014}).

Thus, the recent proposal~\cite{Saha2012} to employ current in the plane of sub-10-nm wide metallic GNRs with zigzag (ZGNRs)~\cite{Jia2009} or chiral (CGNRs)~\cite{Tao2011} edges  hosting  a nanopore  for DNA sequencing has appealing features like large signal (current reaching $\simeq 1$ $\mu$A) at low bias voltage ($\lesssim 0.1$ V), which can change up to 40\% when different nucleobase pass through the nanopore. This is largely driven by peculiar edge electronic currents in ZGNRs and CGNRs, as observed in quantum transport simulations~\cite{Chang2012} when Fermi energy $E_F$ is sufficiently lose to charge neutrality (or Dirac) point (CNP) and confirmed experimentally~\cite{Jia2009}.

However, subsequent studies~\cite{Avdoshenko2013} have found that although the conductance of undoped ZGNRs can be affected by the presence of DNA within the nanopore, it is difficult to distinguish different nucleobases (unless one employs a gate electrode to shift \mbox{$E_F \rightarrow E_F + 0.2$ eV} away from the CNP). Such discrepancies rise an important question about the optimal diameter $D$ of the nanopore (e.g., \mbox{$D \simeq 3.2$ nm} was used in Ref.~\cite{Avdoshenko2013} vs. \mbox{$D \simeq 1.2$ nm} of Ref.~\cite{Saha2012}). Both nanopore and nanogap-based sensors will lose their effectiveness in identifying different nucleobases as their size is increased~\cite{Chen2012}. Another issue, specifically related to ZGNRs or CGNRs, is how sensitive are their edge currents at CNP to the presence of DNA nucleobases, taking into account that such edge currents are not protected by any topological mechanisms (in contrast to, e.g., topological protection of edge currents in GNRs converted into a two-dimensional topological insulator~\cite{Chang2014}). Finally, the above mentioned discrepancies could also originate from the usage of different computational methodologies, which would pose a question about their ability to capture relevant effects.

Here we conduct simulations of electronic transport---using nonequilibrium Green function formalism combined with density functional theory (NEGF+DFT)~\cite{Taylor2001,Brandbyge2002,Areshkin2010} first-principles approach, as well as computationally less expensive semi-empirical
alternatives like NEGF combined with self-consistent charge density functional-based tight-binding (NEGF+SCC-DFTB)~\cite{dftb,Elstner1998,Pecchia2008}
or NEGF combined with self-consistent extended H\"{u}ckel theory (NEGF+SC-EH)~\cite{Stokbro2010,Zahid2005}---in order to quantify the effect of the nucleobase inside the nanopore on total current flowing into the drain electrode or on spatial profiles of local current density around the nanopore of diameter \mbox{$D=1.2$ nm} and \mbox{$D=1.7$ nm} hosted by ZGNRs. The paper is organized as follows. In Sec.~\ref{sec:setup}, we discuss atomistic structure of ZGNR + nanopore setup, where edge carbon atoms of the nanopore are assumed to be passivated by nitrogen~\cite{Saha2012} and edge carbon atoms of ZGNR are passivated by hydrogen. Section~\ref{sec:negfdft} overviews geometry optimization using DFT (based on VASP package~\cite{vasp,Kresse1993,Kresse1996,Kresse1996a}) and first-principles quantum transport simulations using NEGF+DFT framework (based on ATK package~\cite{atk}). In Sec.~\ref{sec:g}, we discuss zero-bias transmission function of ZGNR + nanopore sensor and the corresponding linear-response conductance at room temperature, as well as total current at finite bias voltage. In the same Section, we compare these global electronic transport quantities obtained from NEGF+DFT methodology with those obtained from NEGF+SCC-DFTB and NEGF+SC-EH methodologies (for semi-empirical calculations we use ATK-SE  package~\cite{atk}). Note that NEGF+DFT approach was utilized in Ref.~\cite{Saha2012} while NEGF+SCC-DFTB was utilized in Ref.~\cite{Avdoshenko2013}. The spatial profiles of local current density in the presence and absence of the nucleobase within the nanopore are computed via NEGF+DFT methodology and presented in Sec.~\ref{sec:localcurrent}. We conclude in Sec.~\ref{sec:conclusions}.

\section{Two-terminal ZGNR + nanopore sensor setup}\label{sec:setup}

The two-terminal sensor illustrated in Fig.~\ref{fig:fig1} consist of $N_z$-ZGNR (composed of $N_z$ zigzag chains) as the active region hosting a nanopore,  where ``active'' means that Hamiltonian and GF discussed in Sec.~\ref{sec:negfdft} are computed for this region. The active region is attached to two ideal semi-infinite source (S) and drain (D)  homogeneous $N_z$-ZGNR electrodes of the same width. They are assumed to terminate in macroscopic Fermi liquid reservoirs characterized by the Fermi functions $f_{S,D}(E)$ at electrochemical potentials $\mu_{S,D} = E_F + eV_{S,D}$, so that $V_b = V_S- V_D$ specifies the applied bias voltage. In realistic devices, ZGNR electrodes will eventually need to be connected to metallic electrodes attached to an external battery. However, the fact that GNRs used in experiments are typically rather long and screening takes place over a distance shorter~\cite{Areshkin2010} than the active region justifies the use of semi-infinite ZGNRs as two electrodes in our simulations.

Although ZGNRs are insulating at very low temperatures due to one-dimensional spin-polarized edge states coupled across the width of the nanoribbon, such unusual magnetic ordering and the corresponding band gap is easily destroyed above $\gtrsim 10$ K~\cite{Yazyev2008,Kunstmann2011}. Thus, they can be considered as a good candidate for metallic electrodes and interconnects~\cite{Areshkin2007a}. Close to CNP, electronic current through ZGNRs is confined to flow around their zigzag edges~\cite{Chang2012},  as confirmed in recent experiments~\cite{Jia2009} where such currents were actually utilized to increase the heat dissipation around edge defects and, thereby, rearrange atomic structure locally until sharply defined zigzag edge is achieved.

We consider two different nanopore diameters, \mbox{$D=1.2$ nm} and \mbox{$D=1.7$ nm}, which are drilled in 14-ZGNR of width \mbox{$\approx$ 3.1 nm} or 16-ZGNR of width \mbox{$\approx 3.5$ nm}, respectively. The recent all-atom MD simulations~\cite{Wells2012} of ssDNA strands translocated through graphene nanopores in the presence of solvent have confirmed transport in single nucleotide steps for nanopores of diameter \mbox{$D \gtrsim 1.4$ nm}. We assume that ZGNR edges are passivated by hydrogen, while edge atoms of the nanopore are assumed to covalently bonded to nitrogen.

The focus of our study is to examine microscopic details of how a DNA nucleobase inserted into the nanopore can affect conduction electronic currents in the plane of ZGNR, as well as which Hamiltonian coupled to NEGF properly captures these effects, rather than to attack the ``full problem'' of signal due to different nucleobases and accompanying noise due to (mainly) DNA structural fluctuations under realistic conditions. Because of multiscale nature of the ``full problem''~\cite{Lagerqvist2006,Krems2009}, its handling requires to couple quantum transport simulations to MD simulations supplying the atomic coordinates
of translocated DNA, transmembrane and the solvent. For our simpler task, we choose only thymine (T) nucleobase inside the nanopore in the orientation shown in Fig.~\ref{fig:fig1}, and we also do not consider phosphate and sugar groups of DNA backbone that are always adjacent to each of the nucleobases. This orientation, where the center of mass of T nucleobase (about \mbox{$1.5$ \AA} away from the nanopore wall) and its tilt and rotation angles are the most probable ones (see Fig.~4 in Ref.~\cite{Wells2012}), was extracted from MD trajectories of Ref.~\cite{Wells2012} for poly(dT)$_{20}$ homopolymers translocated through the single layer graphene transmembrane with the bias voltage 500 mV along the direction vertical to graphene [note that in this setup, poly(dT) homopolymer exhibits the greatest number of nucleotide translocation when compared to poly(dA), poly(dC) and poly(dG) homopolymers~\cite{Wells2012}]. Due to strong hydrophobic interaction, the base occupying the nanopore is localized near its boundary, spending very little time in the pore center~\cite{Wells2012}.

\section{Computational methods}\label{sec:negfdft}

Prior to quantum transport simulations discussed below, the active region of the two-terminal device shown in Fig.~\ref{fig:fig1} is first
structurally optimized using VASP package~\cite{vasp,Kresse1993,Kresse1996,Kresse1996a}. The electron-core interactions are described by the projector
augmented wave method~\cite{Blochl1994,Kresse1999}, and we use Perdew-Burke-Ernzerhof~\cite{Perdew1996} parametrization of the generalized gradient
approximation (GGA) for the exchange-correlation (XC) functional. The cutoff energies for the plane wave basis set used to expand the Kohn-Sham (KS) orbitals
are 400 eV for all calculations. A \mbox{$1 \times 1 \times 1$} $k$-point mesh within Monkhorst-Pack scheme is used for the Brillouin zone integration. Structural relaxations and total energy calculations are performed ensuring that the Hellmann-Feynman forces acting on ions are less
than $0.04$ eV/\AA.

\begin{figure*}
\begin{center}
\includegraphics[scale=0.45,angle=0]{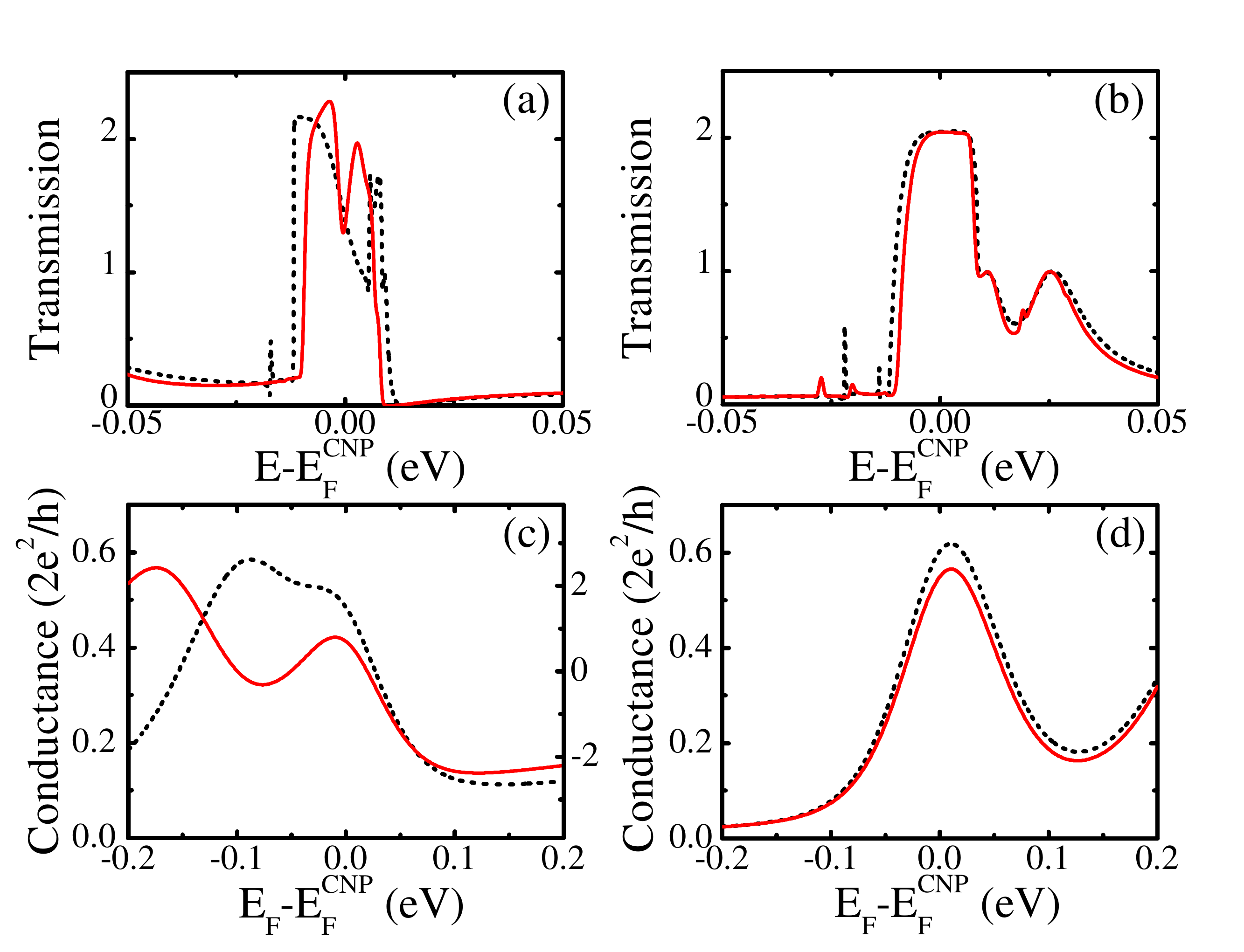}
\end{center}
\caption{The zero-bias transmission function $\mathcal{T}(E)$ in Eq.~\eqref{eq:transzero} for: (a) 14-ZGNR with nanopore of diameter $D=1.2$ nm; and (b) 16-ZGNR with nanopore of diameter $D=1.7$ nm. The conductance at room temperature (\mbox{$T=300$ K}) in panels (c) and (d) is obtained by plugging $\mathcal{T}(E)$ from  panels (a) and (b), respectively, into Eq.~\eqref{eq:conductance}. The dashed line plots electronic transport quantities for ZGNR + empty nanopore, while solid  line plots the same quantities for ZGNR + nanopore + T-nucleobase (see Fig.~\ref{fig:fig1}). The curves plotted in all panels were obtained using NEGF+DFT methodology (Color figure online)}
\label{fig:fig2}
\end{figure*}

In the NEGF+DFT formalism~\cite{Taylor2001,Brandbyge2002,Areshkin2010}, the Hamiltonian of the active region is not known in advance and has to be computed by finding the converged spatial profile of charge via the self-consistent DFT loop for the density matrix \mbox{${\bm \rho} = \frac{1}{2 \pi i} \int dE\, {\bf G}^<(E)$} whose diagonal elements give charge density~\cite{Areshkin2010}. The NEGF formalism~\cite{Stefanucci2013} for steady-state transport operates with two central quantities---retarded ${\bf G}(E)$ and lesser ${\bf G}^<(E)$  GFs---which describe the density of available quantum states and how electrons occupy those states, respectively.

In the coherent transport regime---where we can neglect electron-phonon, electron-electron and electron-spin dephasing processes---only the retarded GF is required to post-process the result of the DFT loop and obtain the zero-bias transmission function between the S and D electrodes as
\begin{equation}\label{eq:transzero}
\mathcal{T}(E) = {\rm Tr} \left\{ {\bm \Gamma}_D (E)  {\bf G}(E) {\bm \Gamma}_S (E)  {\bf G}^\dagger(E)  \right\}.
\end{equation}
The matrices \mbox{${\bm \Gamma}_{S,D}(E)=i[{\bm \Sigma}_{L,R}(E) - {\bm \Sigma}_{S,D}^\dagger(E)]$} describe for the level broadening due to the coupling to the electrodes, where ${\bm \Sigma}_{S,D}(E)$ are the self-energies introduced by the electrodes. The retarded GF matrix of the active device region is given by \mbox{${\bf G}=[E {\bf S} - {\bf H} - {\bm \Sigma}_L - {\bm \Sigma}_R]^{-1}$}, where in the local orbital basis $\{ \phi_i \}$ Hamiltonian matrix ${\bf H}$ is composed of elements \mbox{$H_{ij} = \langle \phi_i |\hat{H}_{\rm KS}| \phi_{j} \rangle$} and $\hat{H}_{\rm KS}$ is the effective KS Hamiltonian obtained from the DFT self-consistent loop. The overlap matrix ${\bf S}$ consists of elements \mbox{$S_{ij} = \langle \phi_i | \phi_j \rangle$}.

\begin{figure}
\begin{center}
\includegraphics[scale=0.3,angle=0]{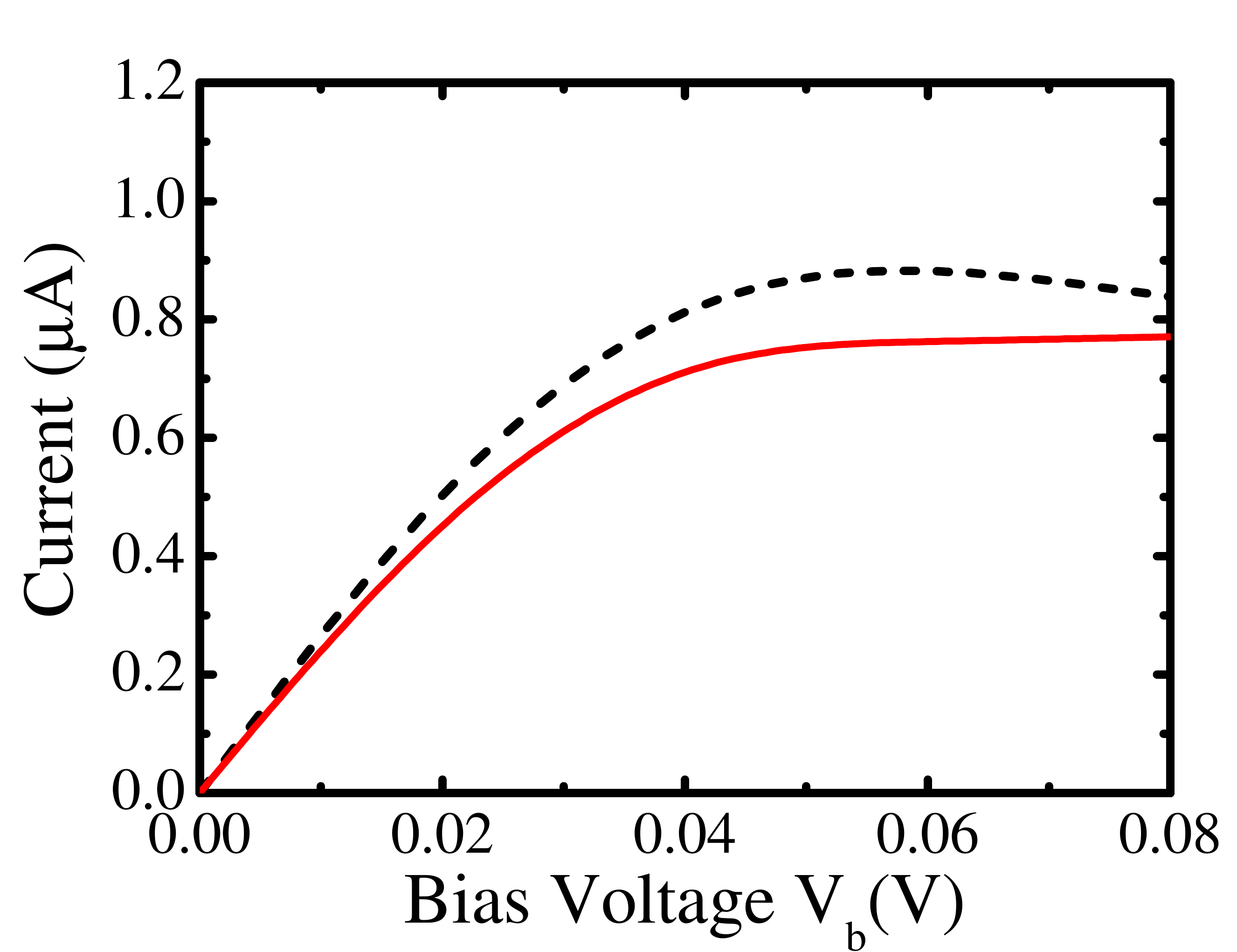}
\end{center}
\caption{The current-voltage characteristics of 16-ZGNR + nanopore (dashed line) or 16-ZGNR + nanopore + T-nucleobase (solid line) systems computed via NEGF+DFT methodology,  where the nanopore diameter is $D=1.7$ nm and T nucleobase is placed within the nanopore in the position depicted in Fig.~\ref{fig:fig1}. The slope of these two curves at vanishingly small bias voltage $V_b \rightarrow 0$ corresponds to conductances in Fig.~\ref{fig:fig2}(d) at $E_F=E_F^\mathrm{CNP}$. The value of the total current on dashed and solid line at $V_b=0.08$ V corresponds to the sum of local currents at an arbitrary cross section in Figs.~\ref{fig:fig7}(a) and ~\ref{fig:fig7}(b), respectively  (Color figure online)}
\label{fig:fig3}
\end{figure}

The linear-response conductance $G=\lim_{V_b \rightarrow 0} I/V_b$ at temperature $T$ is obtained from the transmission function $\mathcal{T}(E)$ using the standard Landauer formula for two-terminal devices~\cite{Stefanucci2013}
\begin{equation}\label{eq:conductance}
G (E_F) = \frac{2e^2}{h} \int\limits_{-\infty}^{+\infty} dE\, {\mathcal T}(E) \left(-\frac{\partial f}{\partial E}\right),
\end{equation}
where $f(E)=\{ 1 + \exp[(E-\mu)/k_BT] \}^{-1}$ is the Fermi function of the macroscopic Fermi liquid reservoirs into which S and D semi-infinite ideal electrodes  terminate. The electrochemical potential $\mu = E_F$ is the same for both reservoirs at vanishingly small bias voltage.

\begin{figure*}
\begin{center}
\includegraphics[scale=0.45,angle=0]{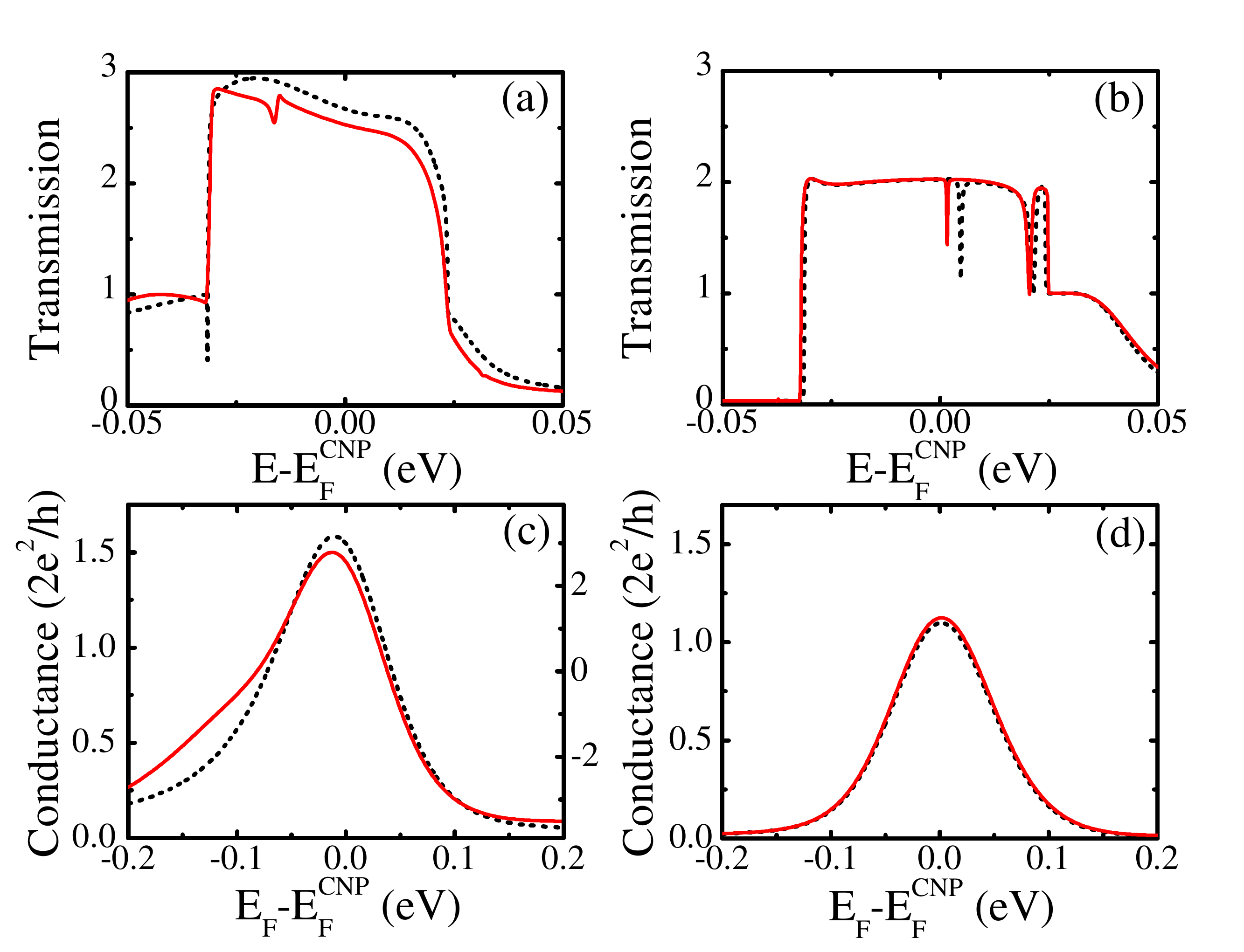}
\end{center}
\caption{The zero-bias transmission function $\mathcal{T}(E)$ in Eq.~\eqref{eq:transzero} for: (a) 14-ZGNR with nanopore of diameter $D=1.2$ nm; and (b) 16-ZGNR with nanopore of diameter $D=1.7$ nm. The conductance at room temperature (\mbox{$T=300$ K}) in panels (c) and (d) is obtained by plugging $\mathcal{T}(E)$ from  panels (a) and (b), respectively, into Eq.~\eqref{eq:conductance}. The dashed line plots electronic transport quantities for ZGNR + empty nanopore, while solid  line plots the same quantities for ZGNR + nanopore + T-nucleobase (see Fig.~\ref{fig:fig1}). The curves plotted in all panels were obtained using NEGF+SCC-DFTB methodology applied to the same device geometries studied in Fig.~\ref{fig:fig2} via NEGF+DFT methodology (Color figure online)}
\label{fig:fig4}
\end{figure*}

When finite bias voltage $V_b=V_S-V_D$ is applied between the source and drain electrodes, we postprocess the result of DFT loop to obtain the
the finite-bias transmission function
\begin{equation}\label{eq:transbias}
\mathcal{T}(E,V_b) = {\rm Tr} \left[ {\bm \Gamma}_D (E,V_D) {\bf G}(E) {\bm \Gamma}_{S}(E,V_S)   {\bf G}^\dagger(E)  \right].
\end{equation}
 Here the self-energy matrices ${\bm \Sigma}_p(E,V_{S,D})={\bm \Sigma}_p(E-eV_{S,D})$ have their electronic band structure rigidly shifted by the applied voltage $eV_{S,D}$. The integration of $\mathcal{T}(E,V_b)$ in Eq.~\eqref{eq:transbias} over the energy window defined by the difference of the Fermi functions
$f_{S,D}(E)=\{1 + \exp[(E-E_F-eV_{S,D})/k_BT]\}^{-1}$ gives the total current
\begin{equation}\label{eq:current}
I = \frac{2e}{h} \int\limits_{-\infty}^{+\infty} dE\, \mathcal{T}(E,V_b) [f_S(E)-f_D(E)],
\end{equation}
flowing through S or D electrode.

The NEGF+DFT simulations of electronic transport are performed using ATK package~\cite{atk} where the local orbital basis $\{ \phi_i \}$ consists of single-zeta polarized pseudoatomic orbitals on C and H atoms and double-zeta polarized on N and O atoms. We use Troullier-Martins norm-conserving  pseudopotentials and  Perdew-Zunger~\cite{Perdew1981} parametrization of the local density approximation (LDA) for the XC functional of DFT. The energy mesh cutoff for the
real-space grid is chosen as 65.0 Hartree. The total number of simulated atoms in the active region, including the T nucleobase within the nanopore, is $657$ in the case of 16-ZGNR and $515$ in the case of 14-ZGNR, respectively.

\begin{figure*}
\begin{center}
\includegraphics[scale=0.45,angle=0]{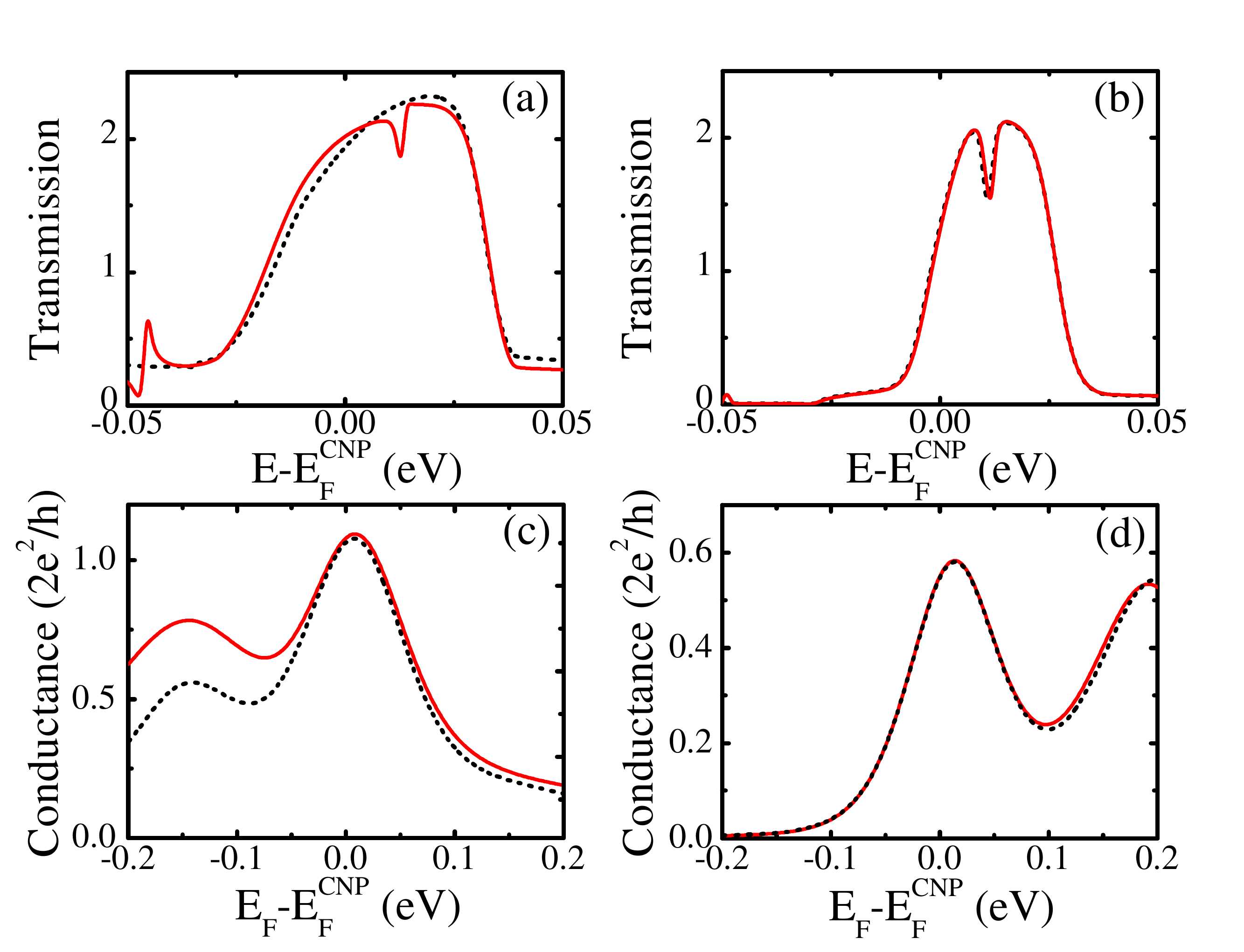}
\end{center}
\caption{The zero-bias transmission function $\mathcal{T}(E)$ in Eq.~\eqref{eq:transzero} for: (a) 14-ZGNR with nanopore of diameter $D=1.2$ nm; and (b) 16-ZGNR with nanopore of diameter $D=1.7$ nm. The conductance at room temperature (\mbox{$T=300$ K}) in panels (c) and (d) is obtained by plugging $\mathcal{T}(E)$ from  panels (a) and (b), respectively, into Eq.~\eqref{eq:conductance}. The dashed line plots electronic transport quantities for ZGNR + empty nanopore, while solid  line plots the same quantities for ZGNR + nanopore + T-nucleobase (see Fig.~\ref{fig:fig1}). The curves plotted in all panels were obtained using NEGF+SC-EH  methodology applied to the same device geometries studied in Fig.~\ref{fig:fig2} via NEGF+DFT methodology (Color figure online)}
\label{fig:fig5}
\end{figure*}

Besides coupling NEGF to DFT Hamiltonian discussed above, computationally much less expensive alternative are offered by semi-empirical methods where electronic
structure is calculated using a model with adjustable parameters fitted to experiments of first-principles calculations. Examples of semi-empirical methods for
electronic transport simulations are based on Slater-Koster tight-binding parameters~\cite{dftb,Elstner1998,Pecchia2008} or EH parameters~\cite{Stokbro2010,Zahid2005}.

We perform NEGF+SCC-DFTB simulations of electronic transport, for the same optimized sensor geometry used in NEGF+DFT-based simulations, via ATK-SE package (which is the semi-empirical part of ATK package~\cite{atk}). In the NEGF+SCC-DFTB methodology~\cite{Pecchia2008}, the non-orthogonal TB-like Hamiltonian $\mathbf{H}$ of the active region, which includes self-consistent potentials, is obtained from SCC-DFTB approach~\cite{Elstner1998} based on second-order expansion of the KS total energy (treated within DFT) with respect to charge density fluctuations. The Slater-Koster parameter file {\tt mio}~\cite{dftb}---developed for organic molecules containing O, N, C, H~\cite{Elstner1998}, S~\cite{Niehaus2001} and P~\cite{Gaus2011} atoms---was employed in SCC-DFTB part of the calculations.

The NEGF+SC-EH simulations are performed using also ATK-SE package~\cite{atk} where a SC Hartree potential is introduced~\cite{Stokbro2010} into conventional EH model in order to take into account the effects of applied bias voltage, external gate potentials or continuum dielectric regions in the device. The details of NEGF+SC-EH implemented in ATK-SE package can be found in Ref.~\cite{Stokbro2010}, where Fermi level of the S and D electrodes is determined self-consistently thereby taking into account (unlike earlier versions of NEGF+SC-EH methodology~\cite{Zahid2005}) the charge transfer from the electrodes to the active region while describing all electrostatic interactions self-consistently. Since properties of graphene largely determine electronic transport through the sensor in
Fig.~\ref{fig:fig1}, we use parameter set for the EH Hamiltonian provided by Ref.~\cite{Cerda2000} which were obtained by fitting the respective bulk band structures.

\section{The effect of DNA nucleobase on total current in the electrodes}\label{sec:g}

\subsection{Results obtained using NEGF+DFT first-principles  methodology}

The zero-bias transmission function $\mathcal{T}(E)$ in Eq.~\eqref{eq:transzero} evaluated for sensors in Fig.~\ref{fig:fig1}, using two different ZGNR widths and nanopores of two different diameters they host, is shown in Figs.~\ref{fig:fig2}(a) and ~\ref{fig:fig2}(b). The corresponding conductances $G(E_F)$ plotted in  Figs.~\ref{fig:fig2}(c) and ~\ref{fig:fig2}(d), respectively, are obtained by integrating  $\mathcal{T}(E)$ in Eq.~\ref{eq:conductance} where we assume that $E_F$ can be shifted by a gate electrode away from the Fermi energy  $E_F^\mathrm{CNP}$ of undoped ZGNRs. At room temperature $T = 300$ K, $-\partial f/\partial E$ is peaked sharply around $E_F$ so that Eq.~\eqref{eq:conductance} depends on the segment of $\mathcal{T}(E)$ curve located within an interval of few $k_BT$ around chosen $E_F$.

The difference in conductances (around $E_F^\mathrm{CNP}$) of empty nanopore and nanopore + T-nucleobase diminishes as the nanopore size is increased from \mbox{$D=1.2$ nm} in Fig.~\ref{fig:fig2}(c) to \mbox{$D=1.7$ nm} Fig.~\ref{fig:fig2}(d). Nevertheless, the contrast for the sensor with the larger nanopore  \mbox{$D=1.7$ nm} can be recovered by applying small bias voltage, as demonstrated in Fig.~\ref{fig:fig3}.

\begin{figure}
\begin{center}
\includegraphics[scale=0.3,angle=0]{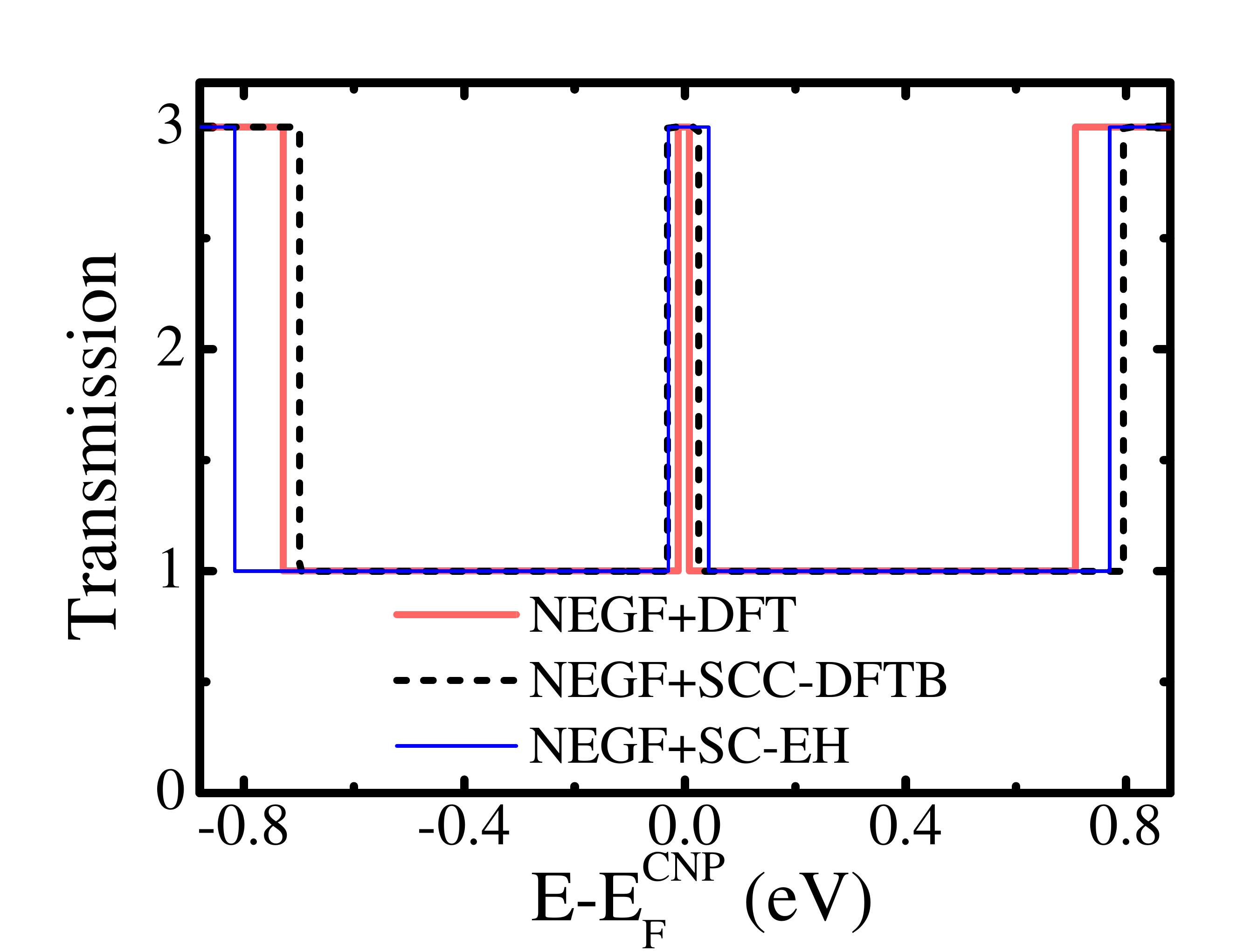}
\end{center}
\caption{The quantized zero-bias transmission function $\mathcal{T}(E)$ in Eq.~\eqref{eq:transzero} for an infinite homogeneous 16-ZGNR computed using NEGF+DFT,  NEGF+SCC-DFTB and NEGF+SC-EH methodologies. The value $\mathcal{T}(E)=3$ around CNP is determined by local edge currents~\cite{Jia2009,Chang2012} in ZGNRs whose spatial profile (for the case when nanopore is drilled within the ZGNR) is depicted in Fig.~\ref{fig:fig7}(a) (Color figure online)}
\label{fig:fig6}
\end{figure}

\subsection{Results obtained using NEGF+SCC-DFTB and NEGF+SC-EH semi-empirical methodologies}

The NEGF+SCC-DFTB methodology is considered to be a much less computationally expensive alternative to NEGF+DFT framework, and it is often employed in simulations of electronic transport through solid-state sensors for DNA sequencing~\cite{Qiu2014,Avdoshenko2013}. For example, this approach, together with other popular~\cite{Lagerqvist2006,Krems2009,Stokbro2010} semi-empirical methods, makes it possible to take snapshots of the atomic coordinates of ssDNA passing through the nanopore or nanogap from time steps of MD simulations and construct a semi-empirical TB-like Hamiltonian for quantum transport calculations of current using formulas discussed in Sec.~\ref{sec:negfdft}. On the other hand, coupling NEGF+DFT to MD simulations requires to select much smaller number of snapshots along the MD trajectory because of computationally much more expensive calculation of electronic structure for each snapshot via DFT self-consistent loop~\cite{Pemmaraju2010}.

However, comparing NEGF+DFT results obtained in Fig.~\ref{fig:fig2} with NEGF+SCC-DFTB results obtained in Fig.~\ref{fig:fig4} reveals large discrepancy between them.
The discrepancy persists also when comparing NEGF+DFT results in Fig.~\ref{fig:fig2} with NEGF+SC-EH results in Fig.~\ref{fig:fig5}. This can be related to the fact that although all three calculations find correct $\mathcal{T}(E)=3$ in infinite homogeneous ZGNRs around CNP, as shown in Fig.~\ref{fig:fig6} this step is much wider and more robust (as also concluded by Ref.~\cite{Avdoshenko2013}) with respect to drilling the nanopores or inserting DNA into the pore in NEGF+SCC-DFTB or NEGF+SC-EH semi-empirical descriptions. This leads to artifactual conclusion that electronic conductance of both small and larger nanopore is virtually insensitive to the presence of DNA nucleobase when $E_F$ of the ZGNR is close the CNP, as shown in Figs.~\ref{fig:fig4}(c), \ref{fig:fig4}(d), \ref{fig:fig5}(c) and \ref{fig:fig5}(d). We note that completely different NEGF+DFT codes yield almost identical results~\cite{Saha2012} for the conductance of ZGNR + nanopore system in the presence or absence of DNA nucleobases, thereby suggesting that NEGF+DFT methodology captures more reliably charge transfer and charge redistribution when constructing the self-consistent Hamiltonian of such systems.

\begin{figure}
\begin{center}
\includegraphics[scale=0.8,angle=0]{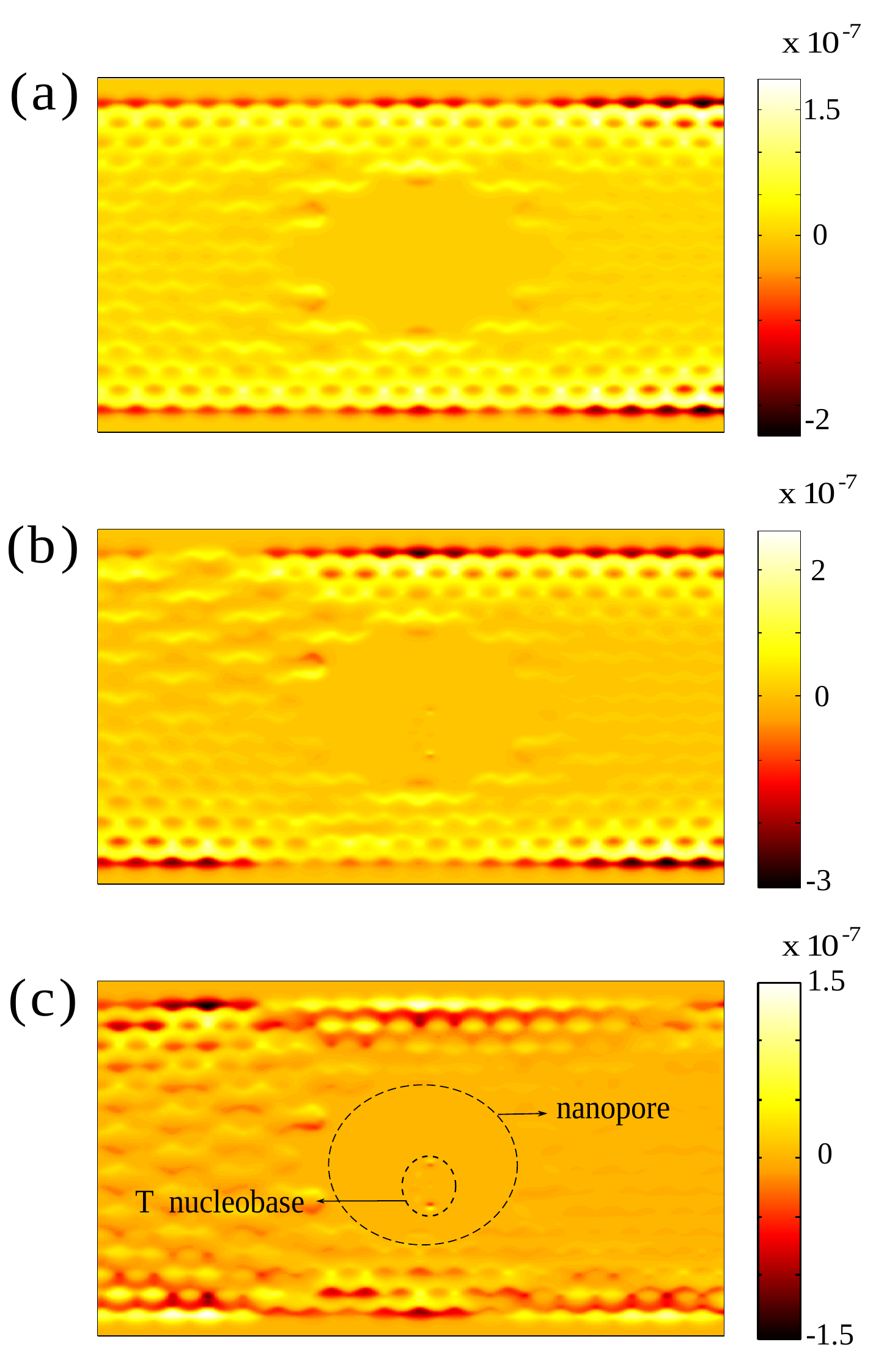}
\end{center}
\caption{The NEGF+DFT-computed spatial profiles of: (a) current density in 16-ZGNR + nanopore; (b) current density in 16-ZGNR + nanopore + T-nucleobase; and (c) difference in current densities between panels (a) and (b). The nanopore diameter is $D=1.7$ nm and T nucleobase is positioned within the nanopore as shown in Fig.~\ref{fig:fig1}. The sum of local current densities in panels (a) and (b) over any vertical cross section gives the corresponding total current at bias voltage $V_b=0.08$ V in Fig.~\ref{fig:fig3} (Color figure online)}
\label{fig:fig7}
\end{figure}

\section{The effect of DNA nucleobase on spatial profiles of current density around the nanopore}\label{sec:localcurrent}

The nucleobase inserted into the nanopore can in principle affect either the local edge currents in ZGNR via modification of the electrostatic potential around the nanopore~\cite{Nelson2010,Saha2012,Xie2012}, or it can induce additional current density around the nanopore wall. To clarify which of these
microscopic mechanisms are contributing to the change of total current shown in Fig.~\ref{fig:fig3}, we compute spatial profiles of local current density
within ZGNR hosting \mbox{$D=1.7$ nm} nanopore.

The knowledge of the lesser GF, $\mathbf{G}^<(E) = \mathbf{G}(E) \cdot {\bm \Sigma}^<(E) \cdot \mathbf{G}^\dagger(E)$, makes it possible to compute the current density using~\cite{Stefanucci2013}
\begin{equation}\label{eq:currentdensity}
\mathbf{J}(\mathbf{r}) = - \frac{e \hbar}{4 \pi m} \int dE \, [(\nabla - \nabla^\prime) G^< (\mathbf{r},\mathbf{r}^\prime;E)]_{\mathbf{r}=\mathbf{r}^\prime}.
\end{equation}
Here the lesser self-energy is given by ${\bm \Sigma}^<(E) = i \sum_{\alpha=S,D} f_\alpha(E) {\bm \Gamma}_\alpha(E)$. The real-space representation of the lesser GF can be obtained from its representation in the local orbital basis $\{ \phi_i(\mathbf{r})\}$ using
\begin{equation}\label{eq:lesser}
G^<(\mathbf{r},\mathbf{r}^\prime;E) = \sum_{i,j} \phi_i(\mathbf{r}) G^<_{ij}(E) \phi^*_j(\mathbf{r}^\prime),
\end{equation}
where $i,j$ sum over all orbitals in the active region and $G^<_{ij}(E)$ denote the matrix elements in the orbital space.

The computation~\cite{Zhang2011b} of $\mathbf{J}(\mathbf{r})$ in Eq.~\eqref{eq:currentdensity} using NEGF+DFT framework applied to 16-ZGNR + nanopore and 16-ZGNR + nanopore + T-nucleobase systems at the applied bias voltage \mbox{$V_b =0.08$ V} leads to spatial profiles shown in Figs.~\ref{fig:fig7}(a) and ~\ref{fig:fig7}(b), respectively. For clarity, we also plot the difference between current densities in these two spatial profiles in Fig.~\ref{fig:fig7}(c). The profiles in Figs.~\ref{fig:fig7}(b) and ~\ref{fig:fig7}(c) demonstrate visible change in local currents flowing along the zigzag edges, as well as introduction of non-zero current density at the position of the T nucleobase due to its close proximity to the nanopore wall. This then explains the difference of the corresponding total currents plotted in Fig.~\ref{fig:fig3} at the same bias voltage.

\section{Conclusions} \label{sec:conclusions}
In conclusion, by comparing first-principles NEGF+DFT with semi-empirical NEGF+SCC-DFTB and NEGF+SC-EH simulations of electronic transport through ZGNR + nanopore sensors, we reexamine recent proposal~\cite{Saha2012} to employ their transverse conduction (rather than usually considered tunneling~\cite{Zwolak2008,Lagerqvist2006,Krems2009,Meunier2008,Chen2012,Tsutsui2010,Huang2010}) current of the order of \mbox{$\sim 1$ $\mu$A} for rapid sequencing of DNA translocated through the nanopore. Contrary to the conclusions based on NEGF+SCC-DFTB (see also Ref.~\cite{Avdoshenko2013}) or NEGF+SC-EH simulations, which find (see Figs.~\ref{fig:fig4} and \ref{fig:fig5}) small effect of DNA nucleobases inserted into the nanopore on the ``robust'' electronic currents flowing along the edges of ZGNRs (for $E_F$ close to CNP), NEGF+DFT simulations show large effect in small nanopores which diminishes with increasing nanopore diameter. However, by applying small bias voltage \mbox{$V_b \lesssim 0.1$ V} the sensitivity can be recovered even for larger nanopores. Besides clarifying the  choice of computational methodology, we also mention that discrepancies generated by prior NEGF+DFT~\cite{Saha2012} and NEGF+SCC-DFTB studies~\cite{Avdoshenko2013} were partly due to positioning of nucleobases within the nanopore in orientations that are highly improbable according to MD trajectories~\cite{Wells2012}. That is, in both studies  nucleobases were placed in the center of the nanopore, rather than close to the nanopore wall as in Fig.~\ref{fig:fig1}, with the plane of the nucleobase orthogonal~\cite{Saha2012} to graphene plane ($\beta =90^\circ$ in Fig.~\ref{fig:fig1}) or aligned~\cite{Avdoshenko2013}  with graphene plane ($\beta=0^\circ$ in Fig.~\ref{fig:fig1}). Our study suggests that by using information about the clustering of nucleobase orientations within the graphene nanopore extracted from MD simulations~\cite{Wells2012}, and by computing spatial profiles of current density for those orientations via NEGF+DFT simulations, one can develop efficient algorithms to search for optimal geometry of GNR + nanopore sensors for rapid DNA sequencing.

\begin{acknowledgements}
We thank A. Aksimentiev for illuminating discussions and for providing us with MD simulation data from Ref.~\cite{Wells2012}, as well as K. K. Saha for technical help. P.-H. C. and B. K. N. were supported by NSF under Grant No. ECCS 1202069. H. L. was supported by NSFC under Grant No. 21303072 and the Science Foundation for Middle-aged and Young Scientist of Shandong Province under Grant No. BS2012DX002. The supercomputing time was provided in part by NSF through XSEDE resource TACC Stampede.
\end{acknowledgements}

%\bibliographystyle{IEEEtran}
%\bibliography{D:/PHYSICS/TEX/BIBTEX/qttg}

\begin{thebibliography}{10}
\providecommand{\url}[1]{#1}
\csname url@samestyle\endcsname
\providecommand{\newblock}{\relax}
\providecommand{\bibinfo}[2]{#2}
\providecommand{\BIBentrySTDinterwordspacing}{\spaceskip=0pt\relax}
\providecommand{\BIBentryALTinterwordstretchfactor}{4}
\providecommand{\BIBentryALTinterwordspacing}{\spaceskip=\fontdimen2\font plus
\BIBentryALTinterwordstretchfactor\fontdimen3\font minus
  \fontdimen4\font\relax}
\providecommand{\BIBforeignlanguage}[2]{{%
\expandafter\ifx\csname l@#1\endcsname\relax
\typeout{** WARNING: IEEEtran.bst: No hyphenation pattern has been}%
\typeout{** loaded for the language `#1'. Using the pattern for}%
\typeout{** the default language instead.}%
\else
\language=\csname l@#1\endcsname
\fi
#2}}
\providecommand{\BIBdecl}{\relax}
\BIBdecl

\bibitem{Schadt2010}
E.~E. Schadt, S.~Turner, and A.~Kasarskis, A window into third-generation sequencing, \emph{Human Molecular Genetics} {\bf 19}, R227 (2010)

\bibitem{Venkatesan2011}
B.~M. Venkatesan and R.~Bashir, Nanopore sensors for nucleic acid analysis,  \emph{Nature Nanotech.} {\bf 6}, 615 (2011)

\bibitem{Wanunu2012}
M.~Wanunu, Nanopores: A journey towards DNA sequencing, \emph{Physics of Life Reviews} {\bf 9}, 125 (2012)

\bibitem{Dekker2007}
C.~Dekker, Solid-state nanopores, \emph{Nature Nanotech.} {\bf 2}, 209 (2007)

\bibitem{McNally2010}
B.~McNally, A.~Singer, Z.~Yu, Y.~Sun, Z.~Weng, and A.~Meller, Optical
  recognition of converted DNA nucleotides for single-molecule DNA sequencing
  using nanopore arrays, \emph{Nano Lett.} {\bf 10}, 2237 (2010)

\bibitem{Merchant2010}
C.~A. Merchant, K.~Healy, M.~Wanunu, V.~Ray, N.~Peterman, J.~Bartel, M.~D.
  Fischbein, K.~Venta, Z.~Luo, A.~T.~C. Johnson, and M.~Drndi\'{c}, DNA
  translocation through graphene nanopores, \emph{Nano Lett.} {\bf 10}, 2915 (2010)

\bibitem{Schneider2010}
G.~F. Schneider, S.~W. Kowalczyk, V.~E. Calado, G.~Pandraud, H.~W. Zandbergen,
  L.~M.~K. Vandersypen, and C.~Dekker, DNA translocation through graphene
  nanopores, \emph{Nano Lett.} {\bf 10}, 3163 (2010)

\bibitem{Garaj2010}
\BIBentryALTinterwordspacing
S.~Garaj, W.~Hubbard, A.~Reina, J.~Kong, D.~Branton, and J.~A. Golovchenko,
Graphene as a subnanometre trans-electrode membrane, \emph{Nature} {\bf 467}, 190 (2010)

\bibitem{Traversi2013}
F.~Traversi, C.~Raillon, S.~M. Benameur, K.~Liu, S.~Khlybov, M.~Tosun,
  D.~Krasnozhon, A.~Kis, and A.~Radenovic, Detecting the translocation of DNA
  through a nanopore using graphene nanoribbons, \emph{Nature
  Nanotech.} {\bf 8}, 939 (2013)

\bibitem{Geim2009}
A.~K. Geim, Graphene: Status and prospects, \emph{Science} {\bf 324}, 1530 (2009)


\bibitem{Schneider2013}
G.~F. Schneider, Q.~Xu, S.~Hage, S.~Luik, J.~N.~H. Spoor, S.~Malladi,
  H.~Zandbergen, and C.~Dekker, Tailoring the hydrophobicity of graphene for
  its use as nanopores for DNA translocation, \emph{Nature Commun.} {\bf 4}, 2619 (2013)

\bibitem{Wells2012}
D.~B. Wells, M.~Belkin, J.~Comer, and A.~Aksimentiev, Assessing graphene
  nanopores for sequencing DNA, \emph{Nano Lett.} {\bf 12}, 4117 (2012)

\bibitem{DasSarma2011}
S.~Das~Sarma, S.~Adam, E.~H. Hwang, and E.~Rossi, Electronic transport in
  two-dimensional graphene, \emph{Rev. Mod. Phys.} {\bf 83}, 407 (2011)

\bibitem{Rocha2013}
A.~R. Rocha, Graphene-based DNA sequencing devices and the effect of the
  environment: A NEGF/QMMM hybrid study, \url{http://www.nordita.org/docs/agenda/dna2013/slides-dna2013-reilyrocha.pdf} (2013)


\bibitem{Zwolak2008}
M.~Zwolak and M.~Di~Ventra, Colloquium: Physical approaches to DNA sequencing
  and detection, \emph{Rev. Mod. Phys.} {\bf 80}, 141 (2008)

\bibitem{Lagerqvist2006}
J.~Lagerqvist, M.~Zwolak, and M.~D. Ventra, Fast DNA sequencing via
  transverse electronic transport, \emph{Nano Lett.} {\bf 6}, 779 (2006)

\bibitem{Krems2009}
M.~Krems, M.~Zwolak, Y.~V. Pershin, and M.~D. Ventra, Effect of noise on DNA
  sequencing via transverse electronic transport, \emph{Biophys. J.} {\bf 97}, 990 (2009)

\bibitem{Meunier2008}
V.~Meunier and P.~S. Krsti\'{c}, Enhancement of the transverse conductance in
  DNA nucleotides, \emph{J. Chem. Phys.} {\bf 128}, 041103 (2008)

\bibitem{Chen2012}
X.~Chen, I.~Rungger, C.~D. Pemmaraju, U.~Schwingenschl\"ogl, and S.~Sanvito,
  First-principles study of high-conductance dna sequencing with carbon
  nanotube electrodes, \emph{Phys. Rev. B} {\bf 85}, 115436 (2012)


\bibitem{Tsutsui2010}
M.~Tsutsui, M.~Taniguchi, K.~Yokota, and T.~Kawai, Identifying single
  nucleotides by tunnelling current, \emph{Nature Nanotech.} {\bf 5}, 286 (2010)

\bibitem{Huang2010}
S.~Huang, J.~He, S.~Chang, P.~Zhang, F.~Liang, S.~Li, M.~Tuchband, A.~Fuhrmann,
  R.~Ros, and S.~Lindsay, Identifying single bases in a DNA oligomer with
  electron tunnelling, \emph{Nature Nanotech.} {\bf 5}, 868 (2010)

\bibitem{Postma2010}
H.~W.~C. Postma, Rapid sequencing of individual DNA molecules in graphene
  nanogaps, \emph{Nano Lett.} {\bf 10}, 420 (2010)

\bibitem{Prasongkit2011}
J.~Prasongkit, A.~Grigoriev, B.~Pathak, R.~Ahuja, and R.~H. Scheicher,
Transverse conductance of DNA nucleotides in a graphene nanogap from first principles,
\emph{Nano Lett.} {\bf 11}, 1941 (2011)

\bibitem{He2011}
Y.~He, R.~H. Scheicher, A.~Grigoriev, R.~Ahuja, S.~Long, Z.~Huo, and M.~Liu,
  Enhanced DNA sequencing performance through edge-hydrogenation of graphene
  electrodes, \emph{Adv. Funct. Mater.} {\bf 21}, 2674 (2011)

\bibitem{Nelson2010}
T.~Nelson, B.~Zhang, and O.~V. Prezhdo, Detection of nucleic acids with
  graphene nanopores: Ab initio characterization of a novel sequencing
  device, \emph{Nano Lett.} {\bf 10}, 3237 (2010)

\bibitem{Girdhar2013}
A.~Girdhar, C.~Sathe, K.~Schulten, , and J.-P. Leburton, Graphene quantum
  point contact transistor for DNA sensing, PNAS doi: 10.1073/pnas.1308885110 (2013)

\bibitem{Qiu2014}
W.~Qiu, P.~Nguyen, and E.~Skafidas, Graphene nanopores: Electronic transport
  properties and design methodology, \emph{Phys. Chem. Chem. Phys.} {\bf 16}, 1451 (2014)

\bibitem{Chang2010}
S.~Chang, S.~Huang, J.~He, F.~Liang, P.~Zhang, S.~Li, X.~Chen, O.~Sankey, and
  S.~Lindsay, Electronic signatures of all four DNA nucleotides in a
  tunneling gap, \emph{Nano Lett.} {\bf 10}, 1070 (2010)

\bibitem{Ahmed2014}
T.~Ahmed, J.~T. Haraldsen, J.~J. Rehr, M.~D. Ventra, I.~K. Schuller, and A.~V.
  Balatsky, Correlation dynamics and enhanced signals for the identification
  of serial biomolecules and DNA bases, \emph{Nanotech.} {\bf 25}, 125705 (2014)

\bibitem{Saha2012}
K.~K. Saha, M.~Drndi\'{c}, and B.~K. Nikoli\'{c}, DNA base-specific
  modulation of microampere transverse edge currents through a metallic
  graphene nanoribbon with a nanopore, \emph{Nano Lett.} {\bf 12}, 50 (2012)

\bibitem{Jia2009}
X.~Jia, M.~Hofmann, V.~Meunier, B.~G. Sumpter, J.~Campos-Delgado, J.~Manuel,
  R.-H. Hyungbin, S.~Ya-Ping, H.~A. Reina, J.~Kong, M.~Terrones, and M.~S.
  Dresselhaus, Controlled formation of sharp zigzag and armchair edges in
  graphitic nanoribbons, \emph{Science} {\bf 323}, 1701 (2009)

\bibitem{Tao2011}
C.~Tao, L.~Jiao, O.~V. Yazyev, Y.-C. Chen, J.~Feng, X.~Zhang, R.~B. Capaz,
  J.~M. T.~A. Zettl, S.~G. Louie, H.~Dai, and M.~F. Crommie, Spatially
  resolving edge states of chiral graphene nanoribbons, \emph{Nature Phys.} {\bf 7}, 616 (2011)

\bibitem{Chang2012}
P.-H. Chang and B.~K. Nikoli\'{c}, Edge currents and nanopore arrays in
  zigzag and chiral graphene nanoribbons as a route toward high-$ZT$
  thermoelectrics, \emph{Phys. Rev. B} {\bf 86}, 041406(R) (2012)

\bibitem{Avdoshenko2013}
S.~M. Avdoshenko, D.~Nozaki, C.~G. da~Rocha, J.~W. Gonz\'{a}lez, M.~H. Lee,
  R.~Gutierrez, and G.~Cuniberti, Dynamic and electronic transport properties
  of DNA translocation through graphene nanopores, \emph{Nano Lett.} {\bf 13}, 1969 (2013)


\bibitem{Chang2014}
P.-H. Chang, M.~S. Bahramy, N.~Nagaosa, and B.~K. Nikoli\'{c}, Giant thermoelectric effect in graphene-based topological insulators with 
heavy adatoms and nanopores, \emph{Nano Lett.} {\bf 14}, 3779 (2014)


\bibitem{Taylor2001}
J.~{Taylor}, H.~{Guo}, and J.~{Wang}, {\em Ab initio} modeling of quantum
  transport properties of molecular electronic devices, \emph{Phys.
  Rev. B} {\bf 63}, 245407 (2001)

\bibitem{Brandbyge2002}
M.~{Brandbyge}, J.-L. {Mozos}, P.~{Ordej{\'o}n}, J.~{Taylor}, and K.~{Stokbro},
Density-functional method for nonequilibrium electron transport,
  \emph{Phys. Rev. B} {\bf 65}, 165401 (2002)

\bibitem{Areshkin2010}
D.~A. Areshkin and B.~K. Nikoli\'{c}, Electron density and transport in
  top-gated graphene nanoribbon devices: First-principles green function
  algorithms for systems containing a large number of atoms, \emph{Phys. Rev.
  B} {\bf 81}, 155450 (2010)

\bibitem{dftb}
DFTB+: Density Functional based Tight Binding (and more), \url{http://www.dftb-plus.info}; The DFTB Website, \url{http://www.dftb.org}

\bibitem{Elstner1998}
M.~Elstner, D.~Porezag, G.~Jungnickel, J.~Elsner, M.~Haugk, T.~Frauenheim,
  S.~Suhai, and G.~Seifert, Self-consistent-charge density-functional
  tight-binding method for simulations of complex materials properties,
  \emph{Phys. Rev. B} {\bf 58}, 7260 (1998)

\bibitem{Pecchia2008}
A.~Pecchia, G.~Penazzi, L.~Salvucci, and A.~D. Carlo, Non-equilibrium Green's
  functions in density functional tight binding: Method and applications,
  \emph{New J. Phys.} {\bf 10}, 065022 (2008)

\bibitem{Stokbro2010}
K. Stokbro, D. E. Petersen, S. Smidstrup, A. Blom, M. Ipsen, and K. Kaasbjerg, Semiempirical model for nanoscale device simulations,
{\em Phys. Rev. B} {\bf 82}, 075420 (2010)

\bibitem{Zahid2005}
F. Zahid, M. Paulsson, E. Polizzi, A. W. Ghosh, L. Siddiqui, and S. Datta, A self-consistent transport model for molecular conduction based on extended Hückel theory with full three-dimensional electrostatics, {\em J. Chem. Phys.} {\bf 123}, 064707 (2005)

\bibitem{vasp}
\url{http://cms.mpi.univie.ac.at/vasp/}

\bibitem{Kresse1993}
G.~Kresse and J.~Hafner, \textit{Ab initio} molecular dynamics for liquid metals, \emph{Phys. Rev. B} {\bf 47}, 558 (1993)

\bibitem{Kresse1996}
G.~Kresse and J.~Furthm\"uller, Efficient iterative schemes for \textit{ab
  initio} total-energy calculations using a plane-wave basis set, \emph{Phys.
  Rev. B} {\bf 54}, 11169 (1996)

\bibitem{Kresse1996a}
G.~Kresse and J.~Furthm\"{u}llerb, Efficiency of ab-initio total energy
  calculations for metals and semiconductors using a plane-wave basis set,
  \emph{Comput. Mater. Sci.} {\bf 6}, 15 (1996)

\bibitem{atk}
Atomistix Toolkit (ATK) 13.8.1, \url{http://www.quantumwise.com}

\bibitem{Yazyev2008}
O.~V. Yazyev and M.~I. Katsnelson, Magnetic correlations at graphene edges:
  Basis for novel spintronics devices, \emph{Phys. Rev. Lett.} {\bf 100}, 047209 (2008)

\bibitem{Kunstmann2011}
J.~Kunstmann, C.~\"Ozdo\ifmmode~\breve{g}\else \u{g}\fi{}an, A.~Quandt, and
  H.~Fehske, Stability of edge states and edge magnetism in graphene
  nanoribbons, \emph{Phys. Rev. B} {\bf 83}, 045414 (2011)

\bibitem{Areshkin2007a}
D.~Areshkin and C.~White, Building blocks for integrated graphene circuits,
  \emph{Nano Lett.} {\bf 7}, 3253 (2007)

\bibitem{Blochl1994}
P.~E. Bl\"ochl, Projector augmented-wave method, \emph{Phys. Rev. B} {\bf 50}, 17953 (1994)

\bibitem{Kresse1999}
G.~Kresse and D.~Joubert, From ultrasoft pseudopotentials to the projector
  augmented-wave method, \emph{Phys. Rev. B} {\bf 59}, 1758 (1999)

\bibitem{Perdew1996}
J.~P. Perdew, K.~Burke, and M.~Ernzerhof, Generalized gradient approximation
  made simple, \emph{Phys. Rev. Lett.} {\bf 77}, 3865 (1996)

\bibitem{Stefanucci2013}
G.~Stefanucci and R.~van Leeuwen, \emph{Nonequilibrium Many-Body Theory of
  Quantum Systems: A Modern Introduction} (Cambridge University Press, Cambridge, 2013)

\bibitem{Perdew1981}
J.~P. Perdew and A.~Zunger, Self-interaction correction to density-functional
  approximations for many-electron systems, \emph{Phys. Rev. B} {\bf 23}, 5048 (1981)

\bibitem{Niehaus2001}
T.~Niehaus, M.~Elstner, T.~Frauenheim, and S.~Suhai, Application of an
  approximate density-functional method to sulfur containing compounds,
  \emph{J. Mol. Struc. (THEOCHEM)} {\bf 541}, 185 (2001)

\bibitem{Gaus2011}
M.~Gaus, Q.~Cui, and M.~Elstner, DFTB3: Extension of the self-consistent-charge density-functional tight-binding method (SCC-DFTB)
  \emph{J. Chem. Theory Comput.} {\bf 7}, 931 (2011)

\bibitem{Cerda2000}
J. Cerd\'{a} and F. Soria, Accurate and transferable extended H\"{u}ckel-type tight-binding parameter, {\em Phys. Rev. B} {\bf 61}, 7965 (2000)

\bibitem{Pemmaraju2010}
C.~D. Pemmaraju, I.~Rungger, X.~Chen, A.~R. Rocha, and S.~Sanvito, \textit{Ab initio} study of electron transport in dry poly(g)-poly(c) $a$-DNA strands, \emph{Phys. Rev. B} {\bf 82}, 125426 (2010)

\bibitem{Zhang2011b}
L. Zhang, B. Wang, and J. Wang, First-principles calculation of current density in molecular devices, \emph{Phys. Rev. B}  {\bf 84}, 115412 (2011)

\bibitem{Xie2012}
P. Xie,	 Q. Xiong,	 Y. Fang,	Q. Qing, and C. M. Lieber, Local electrical potential detection of DNA by nanowire–nanopore sensors, {\em Nature Nanotech.} {\bf 7}, 119 (2012)

\end{thebibliography}


\end{document}